\documentclass[a4paper]{article}
\pdfoutput=1
\usepackage{hyperref}
\hypersetup{colorlinks,allcolors=black}
\usepackage{lmodern}
\usepackage[T1]{fontenc}
\usepackage[utf8]{inputenc}
\usepackage{amssymb,amsmath}
\usepackage[colorinlistoftodos,prependcaption,textsize=tiny]{todonotes}
\hypersetup{unicode=true,
            pdfborder={0 0 0},
            breaklinks=true}
\usepackage{longtable,booktabs}
\IfFileExists{parskip.sty}{%
\usepackage{parskip}
}{
\setlength{\parindent}{0pt}
\setlength{\parskip}{6pt plus 2pt minus 1pt}
}
\usepackage[yyyymmdd,hhmmss]{datetime}
\usepackage{geometry}
\title{End-to-End Testing of Open-Source Hardware Documentation Developed in Large Collaborations}
\usepackage{authblk} 
\usepackage{hyperref}
\usepackage[
        maxnames=1,
        style=authoryear-comp,
        natbib=true,
        backend=bibtex
]{biblatex}
\usepackage{bibspacing}
\author[1,*]{Melinda Yuan}

\author[1]{Aruna Das} 

\author[1]{Sunny Hu}

\author[1]{Aaroosh Ramadorai}

\author[1]{Imaan Sidhu}

\author[1]{Luke Zerrer} 

\author[2]{Jeremiah Alonzo}

\author[2]{Daniel Jarka}

\author[2]{Antonio Lobaccaro}

\author[2]{Leonardo Lobaccaro}

\author[2]{Raymond Provost}

\author[2]{Alex Zhindon-Romero}

\author[2]{Luca Matone}

\author[3]{Szabolcs Marka} 

\author[1]{Zsuzsa Marka}

\affil[1]{Columbia Astrophysics Laboratory, Columbia University in the City of New York, New York, NY 10027, USA}

\affil[2]{Regis High School, New York, NY 10028, USA}

\affil[3]{Department of Physics, Columbia University in the City of New York, New York, NY 10027, USA}

\affil[*]{Corresponding author: melinda.yuan@columbia.edu}

\begin{document}
\maketitle

\begin{abstract}
Large scientific collaborations, often with hundreds or thousands of members, are an excellent opportunity for a case study in best practices implemented while developing open source hardware. Using a publicly available design of timing equipment for gravitational wave detectors as a case study, we evaluated many facets of the open source hardware development, including practices, awareness, documentation, and longevity. Two diverse student teams, composed of high school and college students, participated in an end-to-end exercise of testing publicly-available documented hardware that originated from more than a decade ago. We found that the primary  value of large collaborations lie in the ability to cultivate teamwork, provide a diverse set of role-models, and explore the possibilities of open hardware development of varying complexities. Learning from the experiences of the student groups, we make constructive recommendations where the open source hardware community can learn from the collaborations and vice versa. \\

\end{abstract}

\begin{longtable}[]{@{}l@{}}
\begin{minipage}[t]{0.97\columnwidth}\raggedright\strut

\subsection{Keywords}\label{h.kdz351yp7g7c}

open source, hardware, large collaborations

\strut\end{minipage}\tabularnewline
\bottomrule
\end{longtable}

\section{(1) Introduction}

{Large international collaborations of scientists explore the frontiers of our knowledge and discover game-changing phenomena that captures the imagination of the public worldwide. Whether investigating our genetic heritage or the collision of enigmatic cosmic objects, hardware technology is used on the bleeding edge of human capabilities, often more akin to art than engineering. That is why we refer to the best of these efforts as \emph{instrument science}. The enormous cost of these decades-long projects are often measured in billions of dollars and hundreds, even thousands of scientists---inevitably fully funded by the international taxpayer. Coordinated financial investment is critical for success in fundamental sciences and it places a welcome burden on hardware developers. As a consequence, there is a desire, or even a requirement, to produce hardware that is fully documented and \textit{open to all}. After all, it was paid for by the people.  

In addition to being an incredible resource for society as a whole, open source hardware can foster collaboration between scientific teams and the general public. Consequently, open source hardware projects can increase accessibility to and interest in science. For that reason, it is important to continuously evaluate open hardware principles to see if they truly support transparency, reproducibility, and understanding. 

In order to explore the practical implications of open hardware developed in large collaborative settings, we worked with undergraduates and high school students with an interest in open-source hardware and design but limited practical experience. The motivated student teams conducted an end-to-end exercise to test the extensive and publicly available documentation written over a decade and a half ago from the design of the Laser Interferometer Gravitational-Wave Observatory (LIGO) Timing System \citep{bartos_2010, sullivan_2023}. 

The LIGO detectors \citep{harry_2010} are part of the global network of interferometers, that includes Virgo \citep{acernese_2015}, GEO600 \citep{dooley_2016, affeldt_2014}, and KAGRA \citep{aso_2013, akutsu_2020}, aiming to observe gravitational waves directly. One hundred years after Albert Einstein predicted the existence of gravitational waves, the first observation was made by the Advanced LIGO detectors in Livingston, Louisiana, and Hanford, Washington. These detectors, while located at the same sites, were an upgrade to the initial LIGO detectors and the culmination of a multi-year team effort of research based on the experience of operating the original detectors for a decade. The Advanced LIGO detectors \citep{abbott_gw} have ten times greater sensitivity and thus observe a thousand times bigger volume of the Universe compared to the initial installation, which significantly increased the likelihood of gravitational wave detection.

Making discoveries requires coordinating within the global gravitational-wave detector network and with other astronomy and astrophysics observatories that can  detect electromagnetic and particle counterparts of gravitational waves and thus provide a complete multi-messenger picture of cosmic events. To support the upgrade of LIGO, the initial LIGO timing system needed to be upgraded as well. 
A new design~\citep{bartos_2010, sullivan_2023} was made that ensures the reliable operation of the detectors and also provides precise timing information of observed gravitational wave events. The new design also aimed to strengthen both the diagnostics capability and the ability to be able to track all synchronization errors.

The National Science Foundation (NSF), which provided funding for the design research, mandated that advanced LIGO documentation, including the timing system, be open to the public. Since open science best practices were not mainstream at the time of the design, we decided to test whether outsiders can really make use of the existing public information efficiently and, if not, what changes need to be made. As the design dates back a decade and a half, we also were able to test whether the documentation can survive large timescales. 

We considered that undergraduate and high school students, new to the fields of open-source science and astrophysics, were the best proxies for outsiders as they would consider everything with fresh eyes. For that reason, we designed an end-to-end exercise in which students gained familiarity with the timing system design by simulating the process of manufacturing a board as well as converting the original design files to modern open source formats. They then followed the testing procedure for the boards as written by the original team  of designers. We conducted surveys of students before and after this process to gauge whether or not the exercise had shifted their viewpoints. The students further reflected on the skills and knowledge they wish they had known prior to the exercise and that educators in academia should know before they introduce students to the field. The high school and undergraduate teams comprised of six students each. Thus, we prioritize qualitative feedback rather than quantitative assessments.

The timing system has successfully provided critical information for over hundred cosmic discoveries detected via their gravitational-wave signatures up to date. From high-school students to faculty, an order of two dozen people worldwide was involved in the timing system project at various stages of design through multiple iterations, testing, manufacturing, installation, maintenance, and remanufacturing over ten years, 2007-2017~\citep{bartos_2010, sullivan_2023}. There are several key differences between the original teams of undergraduates, graduate students, engineers and scientists who worked on the historical design and manufacturing of the timing system over ten years ago and the latest cohort of undergraduate and high school students contributing to this study. The original team's objective was far from assessing and creating open-source hardware. Instead, they were prioritizing LIGO goals and objectives and delivered a robust mission critical system on time and on budget. Domain experts were also more closely involved during the original design process. In this new iteration of the project, scientists took more of an observer's and mentor's role and allowed the undergraduates to explore the documentation independently with a fresh eye from the open-source hardware viewpoint. 

The undergraduate team also remodeled the hardware production process leading up to manufacturing. Since more than a decade passed, they looked for supply chain shortages, cost optimized refinements, and obsolete items. While they did not manufacture any hardware, they obtained quotations from manufacturing firms to get a sense of the feasibility of production as well as the change in price. They also conducted a survey of electronics design software in the open-source context and looked into whether their team can contribute to design changes. For that, they needed a design software that was not behind a paywall; they identified the best option and experimented with it. 

The undergraduates and high school students also conducted in-depth tests. Testing the real boards involved following a step-by-step procedure outlined in the advanced LIGO Timing System documentation by checking parameters such as voltage readings and visual signals. Both the high school and undergraduate teams performed this process on LIGO timing system's Leaf and DuoTone boards \citep{slave_test, duotone_test}. Their experiences learned from this process are further summarized in this paper.

\section{(2) End-to-End Exercise}

\begin{figure}[h!]
        \centering
        \includegraphics[width=0.65\textwidth]{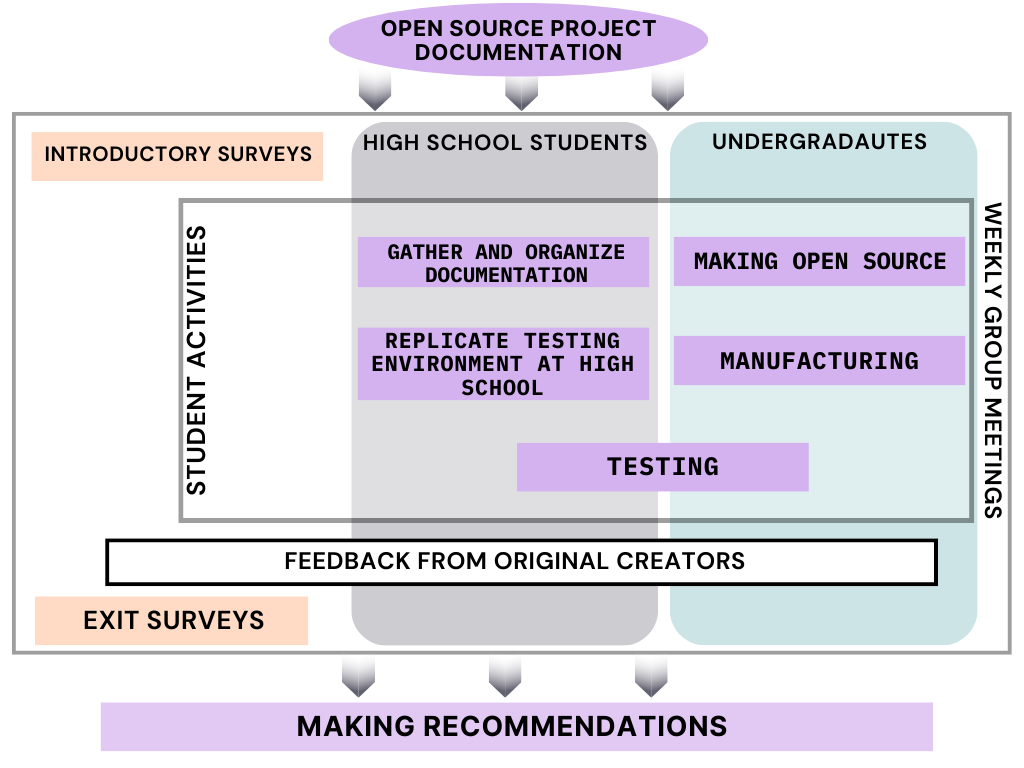}
        \caption{Structure of the end-to-end exercise designed for assessing publicly available documented hardware from an open source point of view.}
\end{figure}
\subsection{Method}

Figure 1 describes the end-to-end exercise designed for assessing publicly available documented hardware from the open source point of view. After introductory team  meetings, all participating students took a survey that assessed  their prior awareness of open-source science in general and open hardware specifically. The high school students were then tasked with collecting and organizing publicly available documentation of the timing system design and manufacturing files through online searches using the public interface of the LIGO Document Control Center \citep{dcc_ligo} 

The undergraduate students were charged with two tasks, which are referred to in this paper as manufacturing and making open source, described below in greater detail. Notably, the students initially had minimal knowledge of both the production and manufacturing of printed circuit boards (PCB) as well as working with PCB design software. 
Subsequently, the student teams were given safety instructions regarding the use of laboratory spaces and equipment, before conducting any tests of previously manufactured boards, furthering their laboratory proficiency.
Finally, at the end of the exercise the students were given an exit survey to assess heir earned experiences.
The end-to-end exercise lasted for approximately half a year during the academic season. In order to facilitate teamwork, the high school students were advised by their physics teacher who has previous experience in LIGO science. The undergraduate team participated in weekly team meetings, and had access to one of the original engineers on the project at an as needed basis. 

\smallskip
\subsection*{Simulating Manufacturing}
The manufacturing group was charged with obtaining quotes from PCB companies for the Leaf board. \footnote{Leaf modules are the terminal points of the timing distribution chain in the LIGO timing system, which has a tree topology. They provide timing information through various parts of the kilometer-scale detector. See \citep{sullivan_2023, slave_test} } By looking at the original PCB design and bill of material files of the Leaf board, they were able to find manufacturers that were (1) domestic, (2) stated that they could provide a full turn key solution, and (3) were ROHS (Restriction of Hazardous Substances) compliant . These companies accepted files in many different formats, the principal three being Altium, Eagle, and KiCad. The acceptance of KiCad is particularly noteworthy as it is one of the only free software suites for electronic design and, thus, the the team evaluated it as the most compatible with open source principles. In the end, the team reached out to around 10 companies and received 7 responses.

The quotes that the team ultimately received were typically divided into the price of bare boards (PCBs without electrical components) and assembly (PCBs that contain all the components) with several different lead times to choose from. Additionally, some manufacturers quoted tooling expenses separately. The group received a total of 5 full quotes, and Figure 2 compares the 2022 prices of these quotes, including the price of a quote from 2017 from the latest remanufacturing run of the same board. Figure 2 also compares the range of lead times from the various manufacturers.

\begin{figure}[h!]
        \centering
        \includegraphics[width=1\textwidth]{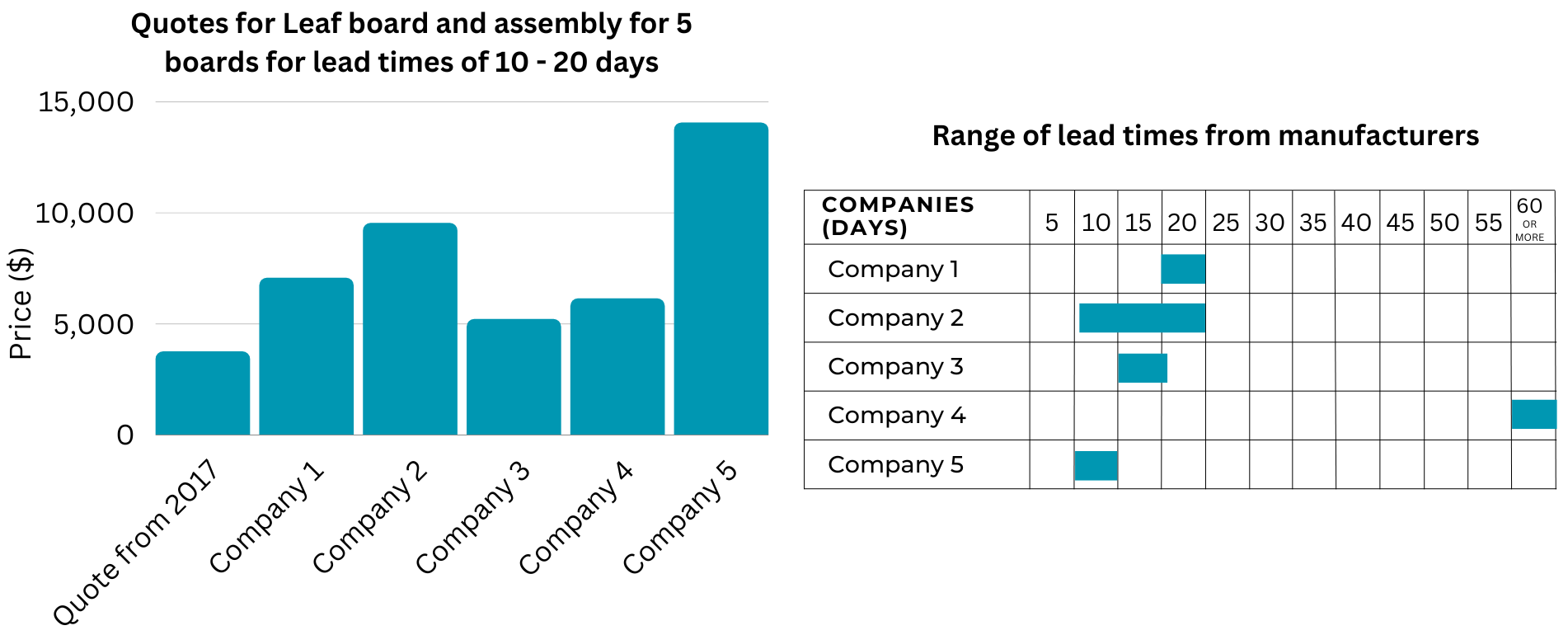}
        \caption{(Left) Price Comparison of Quotes Received. (Please note that the original quote is in 2017 US dollars and the new ones are in 2022 US dollars, and the reader should inflate the 2017 price with about 20\% for proper comparison.) Overall, only one quote approximates the historical price and the rest of the quotes are significantly higher. This might in-part signal that custom hardware manufacturing did not follow the overall inflation models in the US, becoming significantly less affordable for US based creators.}
        \caption{(Right) Range of Lead Times. Please note that while fairly long lead times are acceptable for long term projects like large scientific collaborations, they might represent a significant burden for developers, experimenters, and startups that pride themselves in their agility and speed, either fiercely competing globally or driven by burning enthusiasm.}
\end{figure}

\subsection*{Making Open Source}
The group was supplied with the original PCB files used to manufacture the Leaf Board, which were designed in Altium. The archived format provided was designed using Altium 2009 \citep{altium}. While Altium is among the most popular software for PCB design for professionals, its cost is not conducive to open science or academia. Some engineering students can gain Altium access through their respective institutions, but this is not universal and usually expires by their date of graduation. Further, Altium maybe less available to non-engineering students. Given that the team comprised of both engineering and non-engineering students, they set out to find an open source alternative to Altium for PCB design with sufficient capabilities. 

In determining which open source software to use, the undergraduate team evaluated both price and operating system and determined that KiCad was most suitable software for the project team. KiCad is free, compatible with both Mac and Windows operating systems, and is already in widespread use among engineers and hobbyists \citep{kicad}. The students utilized the excellent tutorials and online resources to become familiar with the KiCad software.

\begin{figure}[h!]
        \centering
        \includegraphics[width=0.6\textwidth]{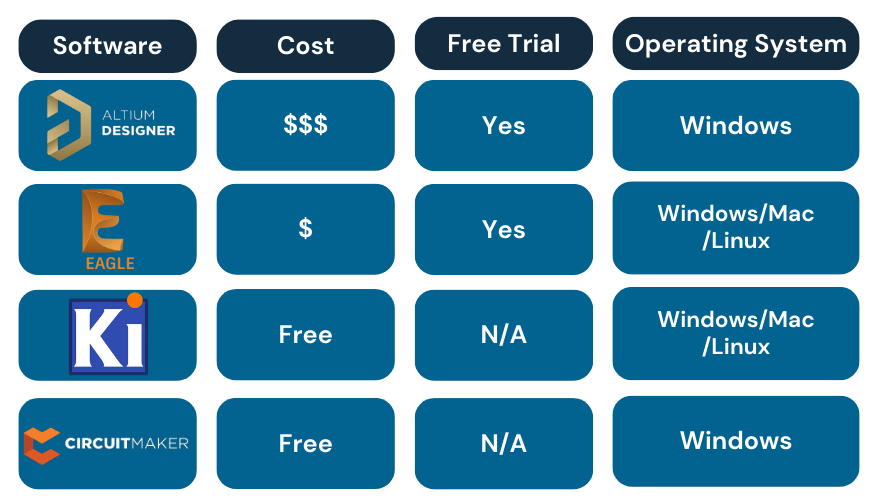}
        \caption{Comparison of Software Options for Electronics Design}
\end{figure}

In their first iteration of the exercise, the students converted the Altium files into KiCad using a third-party tool\citep{altium2kicad}. However, they discovered that the libraries which Altium uses do not match the libraries used in KiCad, rendering the converted file unusable. 

This result led the students to conduct a second iteration of the exercise, in which they eventually settled on a procedure to convert the files manually from Altium to KiCad, which is outlined in Figure 4.

\begin{figure}[h!]
        \centering
        \includegraphics[width=0.7\textwidth]{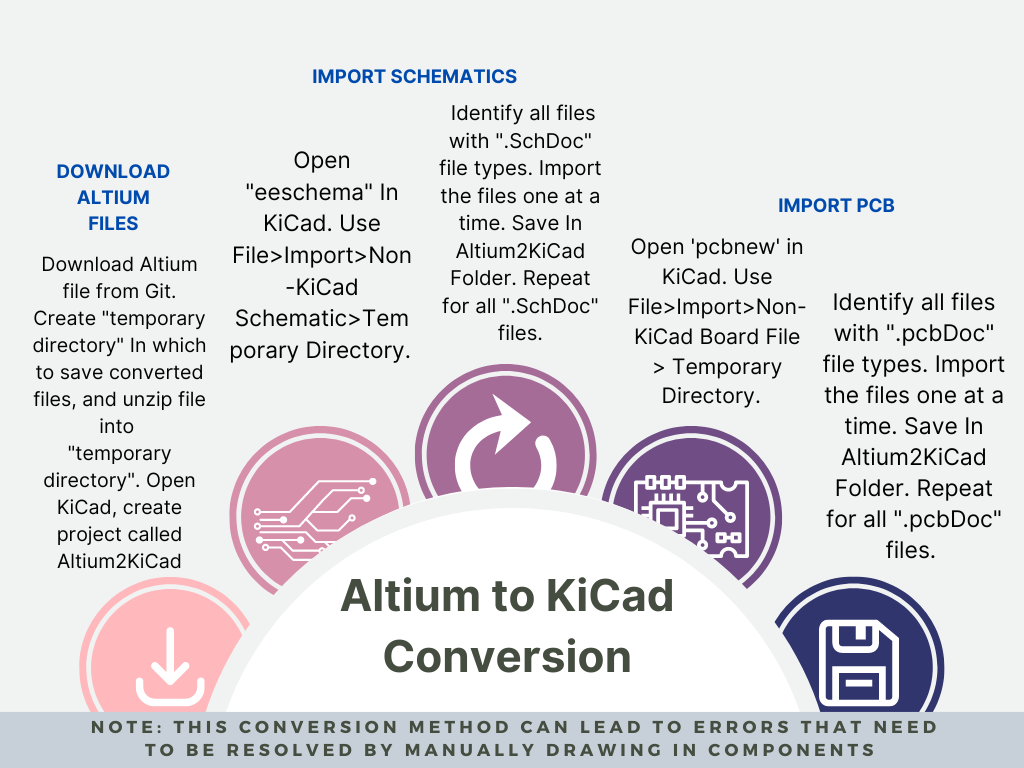}
        \caption{Practical and generally useful process of Altium to KiCad Conversion. Please note the elaborate and time consuming nature of the process.} 
\end{figure}

It should be noted that, while this conversion process was an improvement upon the first attempt, it was also not entirely successful. The gerber files were intact as was most of the PCB layout, but there were several problems with this conversion that required additional manual fixing, including that (1) multi-page schematics were split into multiple schematic documents; (2) there was no link between the schematic files and the PCB layout, which can create problems when attempting to modify; (3) the drill hole sizes of the PCB layout were changed; (4) some wiring was missing in the schematic; and (5) there was no link from the footprints in the schematic to the library.

Once the team had converted the files into KiCad, they were tasked with replacing a custom-made electronics part in the design with a generic version with similar specifications. The custom voltage-control oscillator in the original design was required to fulfill LIGO frequency requirements, but retaining the custom part would have been a barrier to accessibility. Hobbyists or members of the public would not be able to buy this part, and even if they could, custom parts are significantly more expensive than their generic counterparts. Most crucially, the part is not necessary for the broader public; buying the standard available generic part is cheaper, easier, and more useful. 

After selecting the new parts, the students replaced the components on the KiCad schematic, after which they reached out to one of the original designers to verify the compatibility of the substitution.

\subsection*{Testing of Existing Hardware}
The group then set out to test the existing hardware by following the steps from two public LIGO documents: "Test Procedure for the Timing Slave Board" \citep{slave_test} and "Test Procedure for the Clock, Gate, and DuoTone Signal Interface" \citep{duotone_test}. These documents were authored by the members of LIGO who designed, produced, and tested the hardware.

During this process, students encountered a series of challenges, which they carefully documented along with the respective solutions that they devised. These are outlined in the next section.

\section{(3) Challenges and Solutions}

\subsection{Manufacturing Challenges}

\subsubsection{Challenge 1: Old Design and Obsolete Parts}

The fact that the board manufacturing files were originally produced with an older version of Altium from 2009 lead in many cases to additional questions from the production side that needed to be clarified (e.g. drill information), not a simple task for a student new to the field. 
Further, the team had to learn how to handle obsolete parts to which the contacted manufacturers did not have access. As a result, the companies who supplied the team with quotations requested approval of suggested replacement parts for obsolete or out of stock components. This process of replacing obsolete parts then necessitated that the students evaluate the data sheets of the company recommended replacements, posing a myriad of questions. The parts recommended by the companies often had different functionalities from the original parts, making it difficult for the students to determine whether they were suitable replacements. The student group then did further research into the original parts, though information was often limited due to part discontinuation. Changes in certain components would often necessitate further changes in the bill of materials. Due to these issues, the manufacturing group was often unsure of replacement components and, therefore reached out to original designers to confirm their evaluation and the overall production feasibility. 

These aforementioned challenges yielded in some cases difficulties in communication with manufacturers. Not all companies were equally as accommodating with the students, who as stated earlier, were not the original designers, but were doing their first open source hardware project. The team found that smaller companies were usually more responsive, kinder, and willing to provide clarifications and answer additional questions. The larger companies that the students corresponded with were often less willing to provide quotes if they were not guaranteed the order, and were also less accommodating regarding missing information and obsolete parts.

\subsubsection{Takeaway: Contact with Original Creators is Crucial}

In the case of confusion regarding documentation and part replacements, inexperienced individuals should contact those with expertise in engineering and design. The best case scenario, of course, is contact with the original creator of the individual hardware components. The original creator is the most familiar with the design and therefore the most likely to give accurate information. Therefore, when making a hardware project open source, it is crucial for creators to include a way for potential users of the design to contact them, whether it be through email, slack, discord etc. Contact with the original designers or those who keep contributing to the project can increase the longevity of open hardware endeavors. 

\subsection{Testing Challenges}
\subsubsection{Challenge 2: Outdated Legacy Software}
One issue encountered by both teams, but especially by the high school students was working with old versions of software. The testing procedure required test firmware which relied on Altium 2009 and has not been updated since its creation. As a result, the testing of the boards not only required the old version of the software to be installed, but also the old version of the operating system with which the software is compatible with, in this case Windows 7. This required the high school students to find an old computer with the old operating system installed, and then install the old version of Altium. While the students were ultimately able to complete the testing procedure on this old version of the software, the entire process proved not only cumbersome, but would have been virtually impossible without expertise in legacy products which is naturally not the strength of the youngest generation.

\subsubsection{Takeaway: Update Firmware of Design}
The firmware of a design should be regularly updated to keep up with new versions of software that released. Regular updates will maintain the longevity of a design and facilitate use by new users. Alternatively, an install-able snapshot of the original environment and instruction on installation and use should be archived and provided; although it is a poor substitute of a live project.

\subsubsection{Challenge 3: Test Documentation Written for a Tighter Circle of Audience}

Large collaborations consist of people from a multitude of nationalities and education levels — students, professors, engineers, and scientists all over the world — thus there is an effort to make documentation in an accessible fashion for all members. 
In our investigations, both the high school and the undergraduate team encountered difficulties when following some elements of the testing procedure of the boards due to plausible multiple interpretation of text. While documentation was clearly written for scientists who were familiar with the software as well as with the hardware itself, for the truly untrained student, its details were difficult to decipher in several instances. For example, the figures showing the board orientation were not intuitive resulting in difficulty identifying the correct pins for the procedure. Further, he students were unfamiliar with the term "soft LED" and the language of the manual led the students to believe that the LED in question was a physical component on the PCB board, when in fact it was a feature on the Altium screen's control panel.  Such examples show that documentation, obvious to the original design team and even were written with including trained students in the writing, may still have accessibility issues and can be a source of confusion for those who are entirely new to a project. 

\subsubsection{Takeaway: Write clearly for the untrained interested mind of the future}

One of the ways to ensure accessibility is to include students in the writing of documents cataloging the research process. In the long run, comprehensive documentation would allow individuals to contribute to the project without much initial training. Documentation files thus should be written assuming little prior knowledge of the project itself and should be very specific when referencing hardware components. In the process of making a project open source, creators may also benefit from having a team of undergraduates with little prior knowledge about the project test the documentation to determine its true accessibility.

\newpage

\section{(4) Student Recommendations}

The following section contains recommendations derived directly through student feedback in the form of exit surveys. 

\begin{center}

\begin{tabular}{ |p{5cm}|p{8cm}|  }
 \hline
Recommendation & Student Survey Feedback  \\ [1 ex] 
 \hline\hline
Avoid use of legacy software & \textit{“groups that wish to test ... hardware would experience difficulty if they lack access to an expert to clarify confusion or if knowledge of old software”} \\ 
 \hline
 Write detailed and explicit documentation & \textit{“figures showing the board orientation were not intuitive, ... certain indicators of the success or failure of the testing procedure were unclear."}  \\
 \hline
Avoid Obsolescent Parts & \textit{“faulty hardware pieces can also prevent open source hardware from reaching its full potential. However, when maintained properly, open source hardware can be a powerful tool for growing scientists' and the general public's knowledge of the latest hardware.”}

\smallskip
\smallskip
\smallskip

\textit{“a project's design needs to take into account obsolescence and part replacements that have happened since the design was created. Regular maintenance and modification of the design over time to account for this will help overcome this problem.”}

\smallskip
\smallskip
\smallskip

\textit{“precise language in documentation is necessary for understanding instructions years later”}
\\

 \hline
 Be conscious of affordability of parts & \textit{“Price might be a limiting factor. Many manufacturers want larger orders so they charge more per board if you're only ordering a few. The cost of each board is also typically in the hundreds of dollars, which might deter people who are just getting into open source hardware.”}\\
 \hline
 Use old projects as learning tools for future projects & \textit{“Considering part obsolescence and explaining what the function of each part of the design is really important because it allows people to adapt and update your designs. I would consider using old projects for inspiration.”}

\smallskip
\smallskip
\smallskip

\textit{“I would consider using old projects because I see how looking retroactively allowed us to pinpoint exactly what problems to target when it comes to making the project open source.”}
\\ [1ex] 
 \hline
\end{tabular}
\end{center}

\section{(5) Discussion and Additional Recommendations}

The introduction and exit surveys provided valuable insight into the necessary requirements to making a project fully open sourced. Through this end-to-end exercise, our students have provided a myriad of useful recommendations for other educators wishing to pursue an open source hardware project. In addition to those provided by our students, we also collected input from members of LIGO-Virgo-KAGRA, IceCube, and VERITAS Collaborations\footnote{The LIGO Scientific Collaboration (LSC), the Virgo Collaboration and the KAGRA Collaboration, with over 2000 members together,  have joined  to perform gravitational wave science using their respective detectors. The IceCube Neutrino Observatory is a research facility at the South Pole in Antarctica. Over 300 scientists work together in IceCube. VERITAS is a ground-based gamma-ray instrument operating in southern Arizona; the respective collaboration has dozens of members.} on how to make hardware developed in large international collaborative settings more openly sourced. The recommendations made from the aforementioned sources were then compiled into two documents: "Guidelines for Open Source Hardware" and "Mentoring and Training Guide," both of which are available online through the Open Source Hardware Association: https://www.oshwa.org/ (link to documents)

We have highlighted a few key recommendations below:

\begin{itemize}
    \item \textbf{Advocate for Inclusion of Open Source Hardware Standards in Undergraduate Curriculum:} One common theme among both of our student groups was a lack of knowledge about open source hardware in general. Despite being students of the natural sciences, most had never even worked with a PCB board. Regardless of discipline, basic hardware skills are fundamental to science education and require proper advocacy from educators. OSHWA has been actively working to increase awareness of open source hardware. Our documents, which provide guidelines for academic investigators, represent one such effort. Such guidelines are intended to be used by researchers and educators alike to facilitate the incorporation of open source practices in the undergraduate curriculum. For example, the guidelines include practices such as using Git and GitHub for version control, a practice which may be taught in a class or research setting. 
  \item \textbf{Listen to the Student Researchers:} One important takeaway from this exercise is that students know better than anyone else what they know and do not know. Therefore, it is crucial for mentors to listen to the feedback of students, even when it does not necessarily align with their own views. While we often think of a student-mentor relationship as unidirectional, with information flowing from the mentor to the student, the reality is that it is actually a mutual learning process. Especially in the world of open source projects, both the mentor and student are able to learn from each other. 
  \item \textbf{Prioritize Accessibility:} Accessibility requires the consideration of two main aspects: 
 
  (1) Discoverability -- Open source hardware projects must be search engine optimized in a way that is easy to find online. This requires careful planning on the part of the creator. In addition to following open source standards, obtaining a DOI and using key words also contributes to discoverability. 

  (2) Circle of Openness -- The circle of openness refers to the group of individuals who possess the necessary knowledge to be able to access and utilize a project. Deciding on a circle of openness requires consideration of the previous knowledge required to reproduce the hardware, as well as the time investment and learning curve associated. More information regarding the circle of openness is available in Section 1 of "Guidelines for Open Source Hardware." (link) 
  
  The key to the success and longevity of an open source project is accessibility to creators of all backgrounds. Our students experienced difficulties with hardware, software, and testing documentation. While some challenges are bound to arise when embarking on a new project of any kind, creators should strive to anticipate and reduce possible areas of confusion as much as possible. This includes avoiding obsolete parts, updating firmware, and writing precise documentation.
  
  A member of the Virgo collaboration made a suggestion which we thought was worthwhile to mention: to make "simplified" versions of an open hardware project that is intended for the general public. We recommend this approach as another method of increasing the Circle of Openness.
  
  
\end{itemize}

\section{(6) Conclusion}

Our end-to-end exercise proved to be an extremely valuable resource for understanding ways to improve open source hardware. The students were able to gain awareness of the utility of open source hardware as well as its place in the overall open science ecosystem. Moreover, the educators were able to better understand the needs of their students and devise strategies to help other educators incorporate open source hardware into their program. 

While the meticulous design and extensive well-written documentation of the LIGO timing system dates back over a decade and a half, its utility in this exercise sets a precedent for other hardware creators to look not only to the future, but also to the past as a source of inspiration. Only through careful scrutinizing of a previous design was our team able to properly evaluate the longevity of the project. Thus, we encourage other educators and creators alike to constantly look towards past designs and assess their ability to function in a contemporary scientific setting. Only through such explorations of previous work will future projects be improved.

\subsection{Future Work}
Looking towards the future, one change we hope to see is for hardware education to be integrated into the science curriculum. Open software has already made many strides in this area, as open source tools such as Python have been well integrated into the undergraduate curriculum. With the rise of open source hardware fueled by exercises such as the one described in this paper, we hope to see open hardware gain similar traction in terms of awareness and accessibility as open software.

\subsection{Acknowledgements}\label{h.gu3yyarx72d6}

We appreciate the generous support of Open Source Hardware Association (https://www.oshwa.org) and the Sloan Foundation, which awarded Dr. Zsuzsa Márka (Columbia University in the City of New York) the Open Source Hardware Trailblazer Fellowship that made this work possible. Special thanks to Zoltan Raics who was involved as an electrical engineer in the original timing boards. We also thank the LIGO-Virgo-KAGRA, IceCube, and VERITAS collaborations and all of their members. 

\subsection{Funding statement}\label{h.4u1a7tugh2om}

This material is based upon work supported by NSF's LIGO Laboratory which is a major facility fully funded by the National Science Foundation. The authors also gratefully acknowledge support from Columbia University.

\subsection{Competing interests}\label{h.q1j1rznb43fl}
The authors declare that they have no competing interests.

\printbibliography

\end{document}